\documentclass[%
 reprint,
superscriptaddress,
 amsmath,amssymb,
 aps,pre,
]{revtex4-2}

\usepackage{graphicx}
\usepackage{dcolumn}
\usepackage{bm}
\usepackage{color}
\usepackage{times}

\begin{document}

\preprint{APS/123-QED}

\title{Predicting soccer matches with complex networks and machine learning}

\author{Eduardo Alves Baratela}
\affiliation{%
  Instituto de Ciências Matemáticas e de Computação, Universidade de São Paulo, São Carlos 13566-590, Brazil
}
\author{Felipe Jordão Xavier}
\affiliation{%
  Instituto de Ciências Matemáticas e de Computação, Universidade de São Paulo, São Carlos 13566-590, Brazil
}
\author{Thomas Peron}
\email{thomas.peron@usp.br}
\affiliation{%
  Instituto de Ciências Matemáticas e de Computação, Universidade de São Paulo, São Carlos 13566-590, Brazil
}

\author{Paulino Ribeiro Villas-Boas}
\affiliation{%
  Embrapa Instrumentação, Rua XV de Novembro, 1452, São Carlos, SP 13560-970, Brazil
}
\author{Francisco Aparecido Rodrigues}%
\affiliation{%
  Instituto de Ciências Matemáticas e de Computação, Universidade de São Paulo, São Carlos 13566-590, Brazil
}%

\date{\today}

\begin{abstract}
    Soccer attracts the attention of many researchers and professionals in the sports industry. Therefore, the incorporation of science into the sport is constantly growing, with increasing investments in performance analysis and sports prediction industries. This study aims to (i) highlight the use of complex networks as an alternative tool for predicting soccer match outcomes, and (ii) show how the combination of structural analysis of passing networks with match statistical data can provide deeper insights into the game patterns and strategies used by teams. In order to do so, complex network metrics and match statistics were used to build machine learning models that predict the wins and losses of soccer teams in different leagues. The results showed that models based on passing networks were as effective as ``traditional'' models,  which use general match statistics. Another finding was that by combining both approaches, more accurate models were obtained than when they were used separately, demonstrating that the fusion of such approaches can offer a deeper understanding of game patterns, allowing the comprehension of tactics employed by teams relationships between players, their positions, and interactions during matches. It is worth mentioning that both network metrics and match statistics were important and impactful for the mixed model. Furthermore, the use of networks with a lower granularity of temporal evolution (such as creating a network for each half of the match) performed better than a single network for the entire game. 
\end{abstract}

\maketitle


\section{Introduction}

The prediction of soccer match results has been a topic of great interest in the scientific community and the sports industry. Various approaches have been proposed to tackle this problem, typically based on game statistics~\cite{pappalardo2019dataset}, player and team performance analyses~\cite{buldu2018soccer}, and other relevant metrics. However, many of these approaches are limited in their effectiveness and struggle to deal with the complexity and dynamics of player and team interactions.

\textcolor{black}{One promising approach that has gained traction is the application of network science to soccer. This involves the construction of passing networks, which provide a detailed representation of player interactions and game dynamics. Studies have demonstrated that analyzing these networks can yield valuable insights into team strategies and performance~\cite{buldu2019barcelona,gyarmati2014motifs}. For instance, Buldú et al.~\cite{buldu2019barcelona} used network science to analyze Guardiola's FC Barcelona, revealing distinct structural patterns in their passing network that differs from other teams, thus explaining their great performance through network metrics.  Similarly, Gyarmati et al.~\cite{gyarmati2014motifs} introduced the concept of ``flow motifs'' to characterize significant pass sequence patterns in soccer teams. Their analysis demonstrated that, while most teams employ a homogeneous playing style, unique strategies do exist.}

\textcolor{black}{In addition to network science, machine learning has been extensively explored for predicting soccer outcomes. Techniques such as gradient-boosted trees have been employed to learn from relational data and predict match results with significant accuracy~\cite{pappalardo2019dataset}. Machine learning models can leverage a variety of features, including historical match statistics and player performance metrics, to forecast future outcomes~\cite{pena2015Xavi}.}

\textcolor{black}{This paper aims to bridge these two fields by evaluating whether the analysis of passing network structures can enhance the effectiveness of predictive models for soccer match outcomes. To achieve this goal, soccer match data was collected, and passing networks were constructed to analyze player interactions. Complex network metrics were extracted and used as inputs for machine learning models. The effectiveness of these network metrics was compared to traditional models that rely solely on previous game statistics. Furthermore, the combination of network metrics and traditional features was tested to explore potential synergies that could further enhance the predictions of soccer match outcomes. In summary, our results contribute to the field of soccer match outcome prediction by assessing whether the analysis of passing network structures can lead to more effective predictive models. Our findings may provide valuable insights for applying these techniques in other domains and contribute to the improvement of this research field.}

\section{Methodology}

\subsection{ETL / Data Collecting}

     For this research, soccer match data was collected from  \cite{pappalardo2019dataset}. This dataset consists of 1941 soccer matches, including games from the 2018 FIFA World Cup, 2016 UEFA European Championship, and the 2017–2018 seasons of Spanish, Italian, German, English, and French leagues. However, due to the modeling approach used, which considers only wins or losses, games resulting in draws were removed, resulting in a total of 1470 matches. In Appendix~\ref{appendix:appA} we evaluate the performance of our approach when draws are included in the dataset.
    
    
    Each match is recorded in an event-based format, which includes temporal, spatial, and event type information, i.e., the game time and the field coordinates where the events occurred are included. The event data contains detailed information about each interaction between players, where each row in the dataset represents a specific event such as passes, shots on goal, and fouls, including the information of which players performed these events and specific characteristics such as successful or unsuccessful actions, for example, allowing for a more detailed analysis of game dynamics.
    
    This dataset is suitable for the research as it enables the construction of the necessary pass networks to analyze player interactions. Additionally, the variety of teams and competitions included in the data allows for a more robust evaluation of the prediction approaches in a diverse and realistic scenario, including separate tournament-based analyses.

\subsection{Construction of passing networks}

    Passing networks are structures that consider the organization of a team as the result of interaction among its players, thus creating passing networks. These networks are directed (edges between players are unidirectional), weighted (the weight of the edges is based on the number of passes between players), spatial (the Euclidean position of the ball and players is highly relevant), and temporally evolving (the network continuously changes its structure) \cite{buldu2018soccer}.
    
    There are three main types of passing networks, as explained in~\cite{buldu2018soccer}: ``player passing networks'', where each node corresponds to a player on the team; ``pitch passing networks'', where each node corresponds to a delimited region of the field where players occupy; and ``pitch-player passing networks'', which are a combination of the previous two types. In all three approaches, the weight of the edges corresponds to the number of passes between the respective nodes. In this work, we used the first option (player passing networks).
    
    The passing networks were constructed individually for each team, and two different chronological granularities were tested: one network for the entire game and two distinct networks, one for each half of the game. The average field position of each node (player) was also calculated within the given time interval.
    
    It is important to mention that only the initial 22 players of each game were considered in the construction of the networks. The nodes of substitute players were transferred to the players they replaced in order to maintain a consistent structure of 11 nodes for each team in all matches. Although this approach may suppress some team characteristics that change with substitutions, this method was adopted to maintain a consistent structure across all matches, following the methodology proposed by \cite{buldu2019barcelona}.
    
    \begin{figure*}
            \begin{tabular}{cc}
            \includegraphics[scale=0.45]{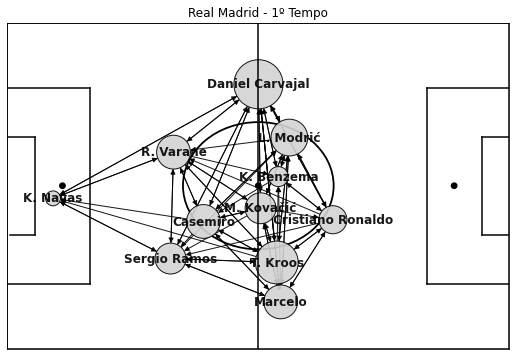} & \includegraphics[scale=0.45]{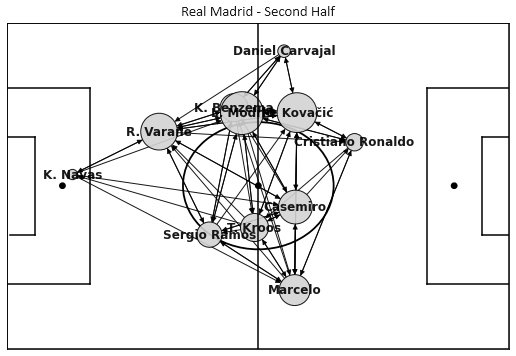} \\
            \includegraphics[scale=0.45]{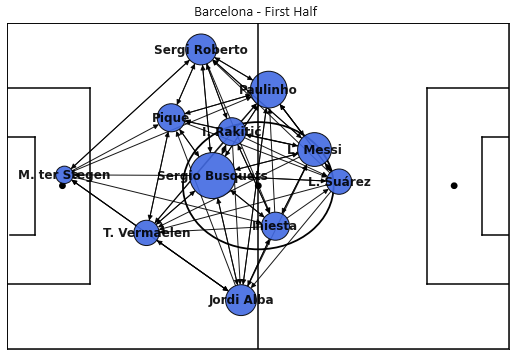} & \includegraphics[scale=0.45]{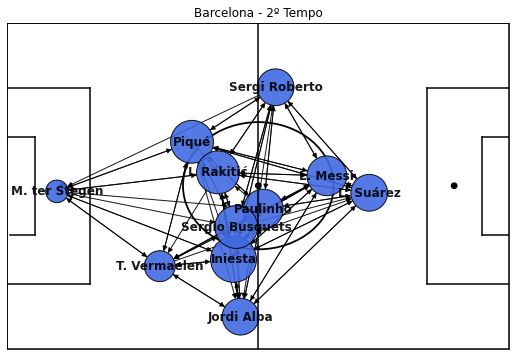} \\
            \end{tabular}
            \caption{Example of a match between Barcelona and Real Madrid, for which four player passing networks were constructed. In the upper part, the networks of Real Madrid for the first and second half (left and right, respectively). In the lower part, the networks of Barcelona for the first and second half, respectively. The larger the area of the node circle, the higher the degree \textcolor{black}{(sum of both in-degree and out-degree)} of that node, indicating more involvement in passes by that player. It is interesting to observe how the networks change from team to team in terms of their structure and priorities on certain players, indicating changes in their characteristics and strategies throughout the game.}
            \label{fig:redes}
        \end{figure*}

\subsection{Measures for network characterization}

    According to \cite{buldu2018soccer}, there are mainly three topological scales in a passing network: microscale, mesoscale, and macroscale.
    
    At the microscale, the importance of each player is assessed individually in the network using metrics such as node degree (the number of passes given and received by each player), eigenvector centrality, closeness centrality, betweenness centrality, clustering coefficient, among others~\cite{buldu2018soccer}.
    
    At the mesoscale, the focus is on analyzing the role of groups of players that are strongly connected to each other. The analysis of network motifs has shown how the abundance of passes between groups of three/four players can be related to both a team's success \cite{gyarmati2014motifs} and the identification of leaders in the passing network \cite{pena2015Xavi}.
    
    Lastly, at the macroscale, the metrics aim to provide information about the style and performance of teams as a whole, such as the position of the network centroid, player dispersion index, and team's average degree.
    
    The following metrics were chosen and used to evaluate the passing networks. For most of these metrics, except for average shortest path and network centroid position, the minimum, maximum, mean, and standard deviation of the nodes were calculated for each network.

\textbf{Degree Centrality}: 
    The measure of a node's importance based on the fraction of nodes it is attached to:
    \begin{equation}
        C_D(v) = \frac{deg(v)}{n-1}, 
    \end{equation}
    where $deg(v)$ is the number of neighbours connected to node $v$, and $n$ is the total number of nodes in the network. In a passing network, $C_D(v)$ indicates the frequency with which a player passes the ball to or receives it from other players.

\textbf{Closeness Centrality}: 
    The measure of how close a node is to all other nodes in the network. It is calculated as the inverse of the sum of the shortest distances from a node to all other nodes:
    \begin{equation}
        C_C(v) = \frac{n-1}{\sum\limits_{u} d(u,v)}, 
    \end{equation}
    where $d(u,v)$ is the shortest distance between nodes $u$ and $v$, measured in number of edges. The closeness centrality indicates how well connected a player is to their teammates, i.e., how many steps the ball has to travel from that player to reach their teammates.

\textbf{Betweenness Centrality}: 
    The measure of how frequently a node appears in all the shortest paths in the network~\cite{newman2018networks}:
    \begin{equation}
         C_B(v) = \sum\limits_{s\neq v\neq t \in V}\frac{\sigma_{st}(v)}{\sigma_{st}}, 
    \end{equation}
    where $V$ is the set of nodes, $\sigma_{st}(v)$ is the number of shortest paths between $s$ and $t$ that go through node $v$, and $\sigma_{st}$ is the total number of shortest paths from node $s$ to $t$.
    In a passing network, betweenness centrality can indicate how important a player is in facilitating communication and coordination within the team.

\textbf{Eigenvector Centrality}: 
    This measure quantifies how well-connected a node is to other nodes in the network, taking into account the importance of those other nodes. Let $\mathbf{A}$ be the adjacency matrix of a network, where the element $A_{ij} = 1$ if there is a link from node $i$ to $j$, whereas $A_{ij}=0$ if this link is absent. The eigenvector centrality of a node $v$, $C_E(v)$, is defined as the $v$-th coordinate of the eigenvector associated with the largest eigenvalue $\lambda_{\max}$ of $\mathbf{A}$~\cite{newman2018networks}, i.e.
    \begin{equation}
    C_E(v) = x(v),\textrm{ where } \mathbf{A}x = \lambda_{\max} x.
    \end{equation}  
    In a passing network, the eigenvector centrality can indicate how influential a player is within the team, i.e., if they pass to other important players in the network.

\textbf{Clustering Coefficient}: 
    The clustering coefficient is the fraction of possible triangles (clusters) that are attached to node $v$. In other words, it measures how well-connected the neighbors of a node are to each other. Let $T(v)$ the number of triangles attached to node $v$. The clustering coefficient is defined as~\cite{newman2018networks} 
       \begin{equation}
    C_{tr}(v) = \frac{2T(v)}{deg(v)(deg(v)-1)}
    \end{equation}
    In a passing network, $C_{tr}(v)$ indicates how likely a player is to pass the ball to other players, who also are likely to pass amongst themselves.

\textbf{Network Centroid Position}: 
    The position of the geometric center of the network, $(x_c,y_c)$. Here we use this metric to quantify the overall positioning and dispersion of the team. Specifically, we will consider the arithmetic mean and the standard deviation of $x_c$ and $y_c$. The $x$ and $y$ axes range from 0 to 100.

\textbf{Average Shortest Path}: 
    The average of the shortest distances between all pairs of nodes in the network, meaning the shortest distance (minimum number of passes) that two distinct players need to connect on the field:
    \begin{equation}
    \ell = \sum\limits_{s,t \in V}\frac{d(s,t)}{n(n-1)}
    \end{equation}
    In a passing network, it indicates how efficiently the team moves the ball across the field.

\subsection{Analysis of team performance}

    The choice of these metrics is based on evidence from various studies \cite{buldu2019barcelona, buldu2018soccer, onody2004brazilian} indicating a correlation between the structure of passing networks and a team's performance on the field, justifying their selection for modeling.

\begin{figure*}
    
        \begin{tabular}{cc}
          \includegraphics[scale=0.43]{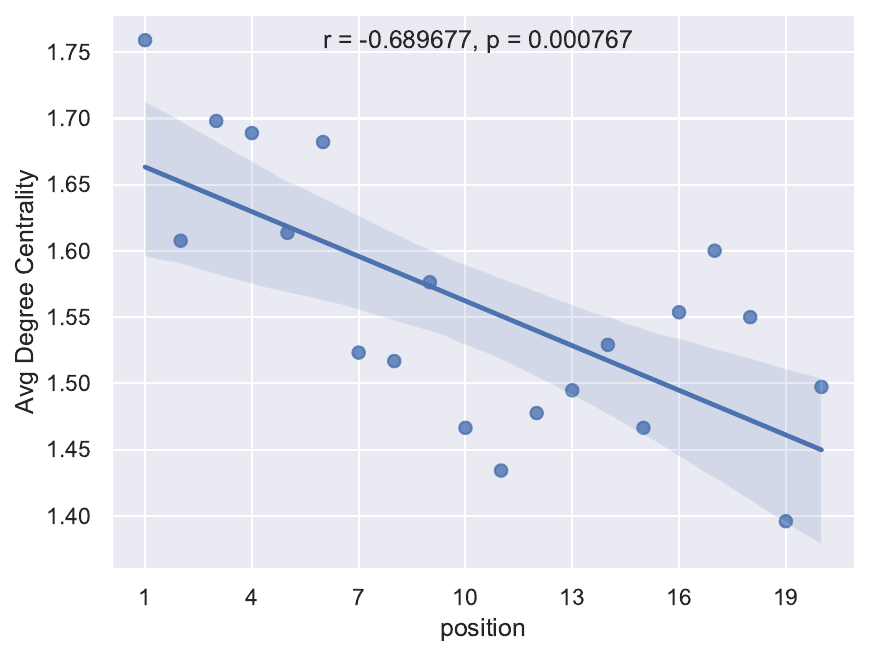} &   \includegraphics[scale=0.43]{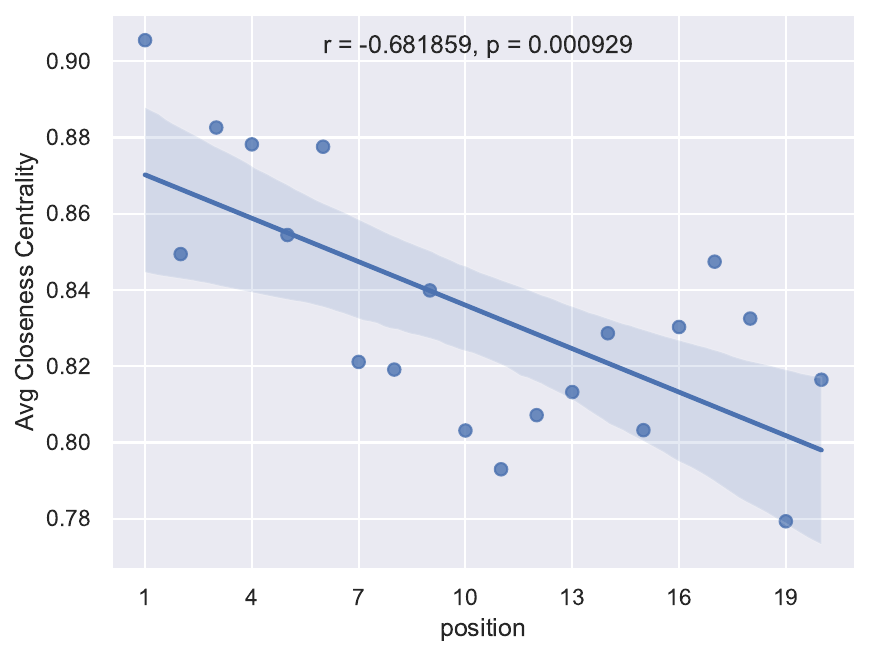} \\
        (a) Degree Centrality & (b) Closeness Centrality \\[6pt]
         \includegraphics[scale=0.43]{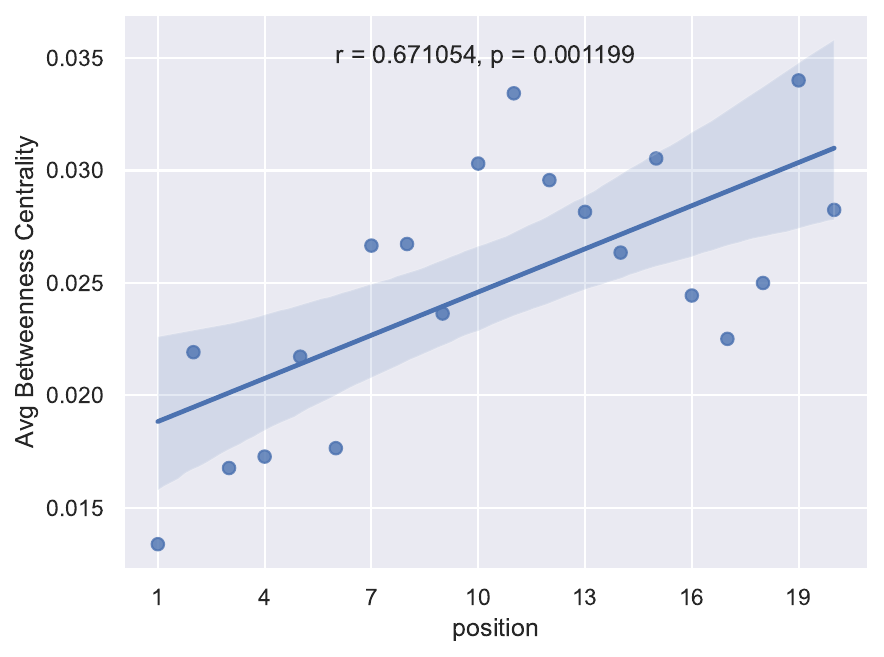} &   \includegraphics[scale=0.43]{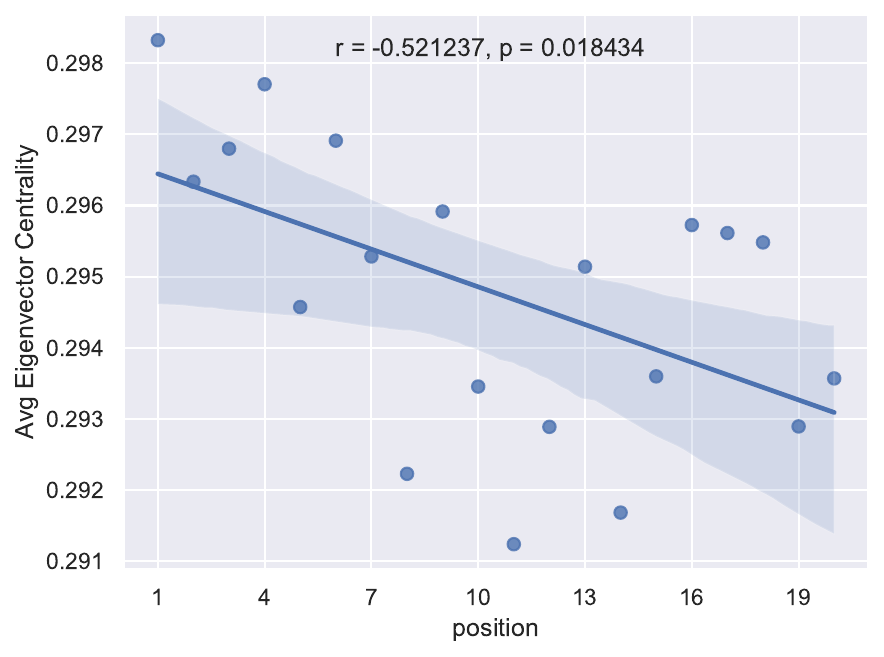} \\
        (c) Betweenness Centrality & (d) Eigenvector Centrality \\[6pt]
        \includegraphics[scale=0.43]{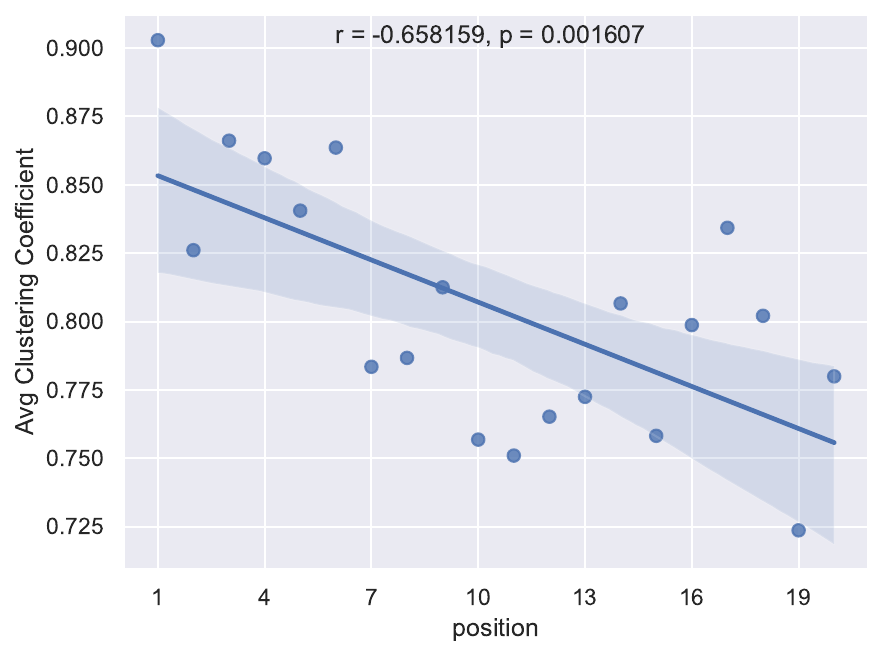} &   \includegraphics[scale=0.43]{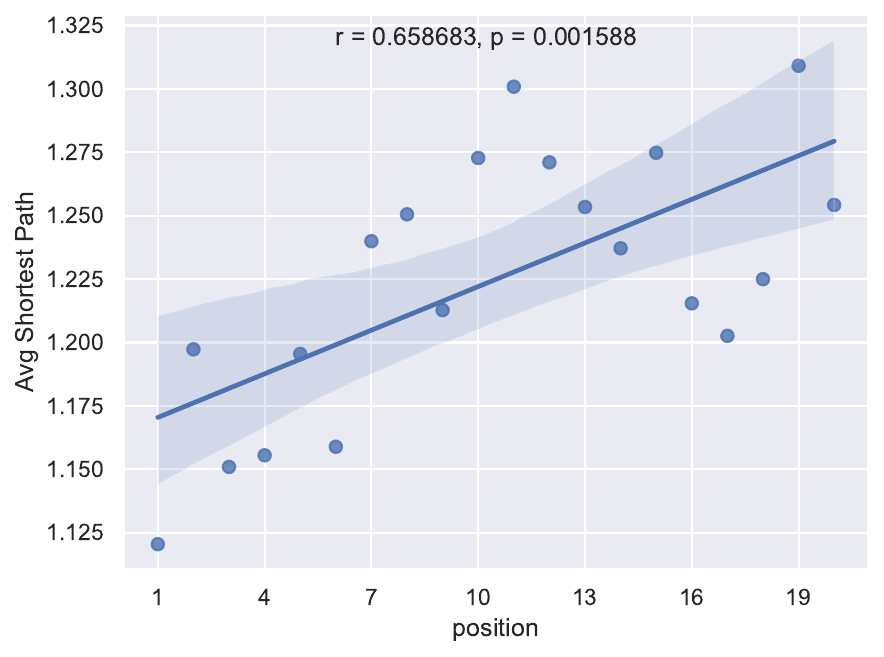} \\
        (e) Clustering Coefficient & (f) Average Shortest Path \\[6pt]
        \includegraphics[scale=0.43]{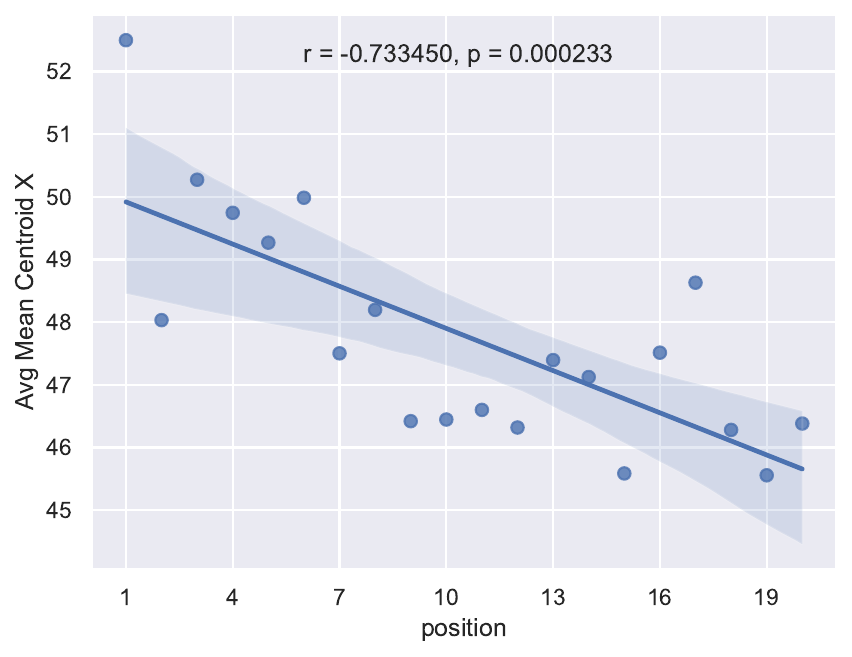} &   \includegraphics[scale=0.43]{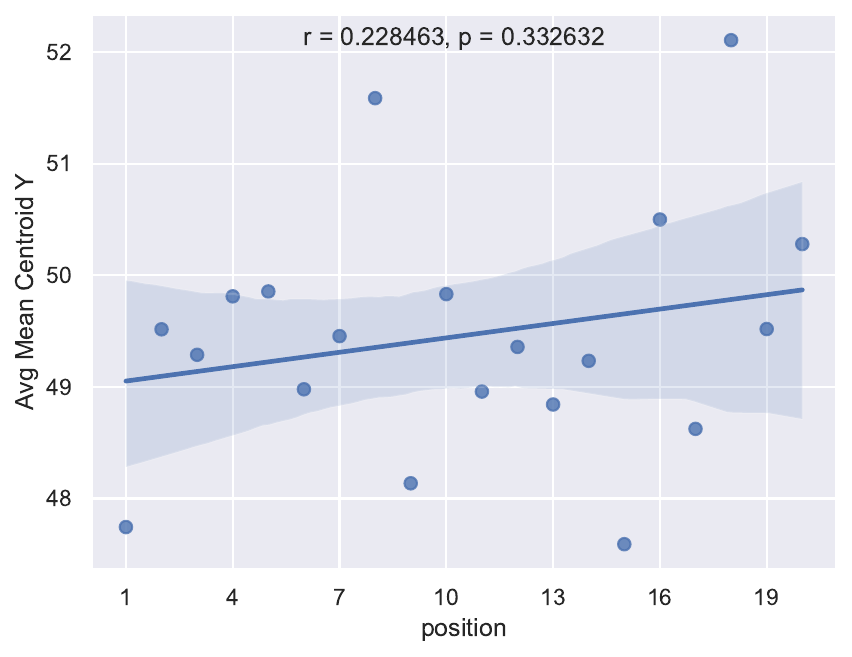} \\
        (g) Network Centroid Position - Eixo X & (h) Network Centroid Position - Eixo Y \\[6pt]
        \end{tabular}
        \caption{Metrics ($y$-axis) in relation to the rankings of Premier League teams ($x$-axis). The values of ``r'' and ``p'' represent the Pearson correlation between the metrics and the teams' rankings in the championship and their respective p-values, respectively.
        \label{fig:metrics}}
    \end{figure*}

    Figure \ref{fig:metrics} presents several graphs showing the correlation between the average of the metrics mentioned earlier and the position of each team in the Premier League (the top division of English soccer). A noteworthy case shown in that figure is the average shortest path. As we seen in panel (f), teams in higher positions in the league tend to have a lower value of $\ell$, indicating an easier connection between two players on the team, requiring fewer passes to create a play. The only metric that does not show any correlation with team performance is the position of the centroid on the Y-axis (Network Centroid Position -- $y$-axis), which is reasonable as it only represents the average lateral positioning of the team. It is also interesting to note that the value for the top-ranked team is always the highest or lowest in all metrics (except for the position of the centroid on the $y$-axis).

\subsection{Machine learning methods}

    In this section, we provide a brief introduction of the machine learning methods used in our study.

    Initially, the $k$-means algorithm was used to perform an exploratory analysis of the structural characteristics of teams in relation to their networks and to check for differences or possible clusters among the leagues of each country. $k$-means \cite{lloyd1982kmeans} is an unsupervised machine learning algorithm that aims to cluster the samples in a dataset into $k$ clusters based on their similarity characteristics. In this case, the similarity between points is measured by the Euclidean distance in the feature space, where each point is assigned to the cluster whose centroid is the closest.

    Subsequently, for predicting the winner of a match, three supervised binary classification models were chosen: Logistic Regression, Random Forest Classifier, and XGBoost.
    
    Logistic Regression \cite{cox1958regression} is a linear model that uses the logistic function to find the relationship between the explanatory variables and the response variable, transforming the model's output into a probability. Random Forest \cite{ho1995random} is a machine learning algorithm that creates multiple decision trees and combines their results to obtain a final prediction. It is an interesting model as it can handle many variables and is less affected by noise in the data. XG Boost (Extreme Gradient Boosting) \cite{Chen:2016:xgboost} is also a model that uses decision trees, but utilizes boosting techniques to create a weighted ensemble of decision trees for prediction. It is a highly popular technique in data science competitions due to its high performance and flexibility \cite{nalluri2020scalable}.
    
    To optimize the models, the Hyperopt library \cite{bergstra2013hyperopt} was used. It is a Python library that utilizes optimization algorithms to find the hyperparameters that maximize the models' performance, potentially leading to improved predictive performance. Additionally, all supervised models were trained using cross-validation, specifically stratified 10-fold cross-validation.
    
    Lastly, to assess important variables and model interpretability, SHAP values were used. SHAP (Shapley Additive Explanations) \cite{lundberg2017shap} is an explainability technique for machine learning models that shows the contribution of each variable to the prediction of a specific outcome.
    
    By using the above methods, we aim to obtain accurate and reliable predictions for match results and understand the characteristics that most influence team performance. It is worth noting that the scikit-learn library \cite{scikit-learn} was used for implementing the predictive models (K-means, Logistic Regression, Random Forest Classifier, and XG Boost).
    
\section{Results}

    The problem we address here is to evaluate the effectiveness of using complex network metrics in machine learning models for predicting soccer match outcomes. The aim is to understand if it is possible to utilize measures that represent the structure of a passing network in a soccer match to predict the winner or if there are other relevant variables that should be considered to obtain more accurate predictions. The objective is to analyze whether the structure of a passing network is a valid indicator for predicting the outcome of a match and how this information can be used in conjunction with other relevant variables, such as match statistics, to improve the accuracy of prediction models.

\subsection{Tournament Clustering}
    
    The clustering of tournaments is a crucial step in understanding whether there are differences in playing styles among different leagues and if there is any possible clustering among them. \textcolor{black}{While it is recognized that soccer is played in all these leagues, the hypothesis is that each league may exhibit unique network structures and match characteristics due to variations in tactical approaches, cultural influences, and historical development of the sport within each country. By analyzing network metrics from tournaments in Spain, Germany, England, France, and Italy, we aim to determine if these differences are significant enough to create distinct clusters.}
    
    \textcolor{black}{This type of analysis is justified as it allows us to explore whether the competitive environment and playing styles are homogeneous within each league or if there are discernible patterns that differentiate leagues from each other. It is important to note that such an analysis also helps to identify if the differences within a league (e.g., between the first and last team) are more pronounced than the differences between leagues. By clustering the tournaments based on network metrics, we can gain insights into the structural and tactical diversity present in European soccer and better understand the global landscape of the sport.}

    \begin{figure*}
        \begin{tabular}{cc}
          \includegraphics[scale=0.44]{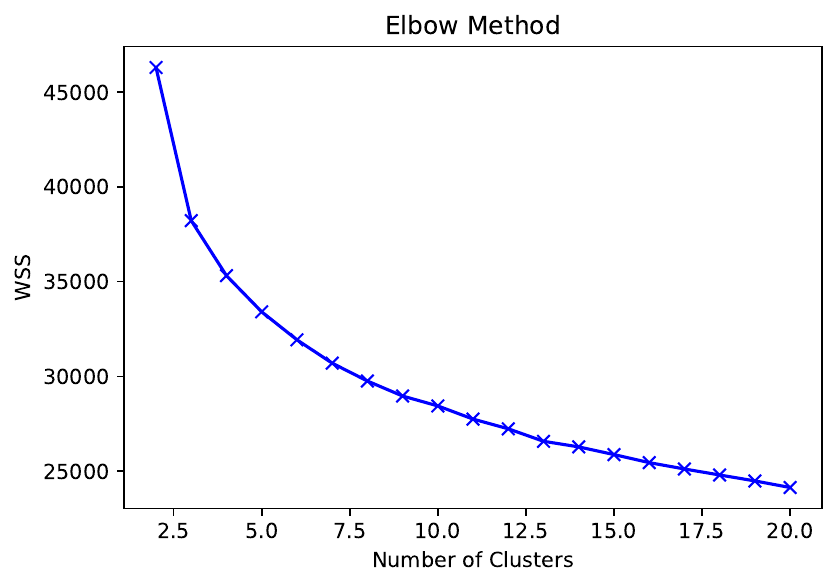} &   \includegraphics[scale=0.38]{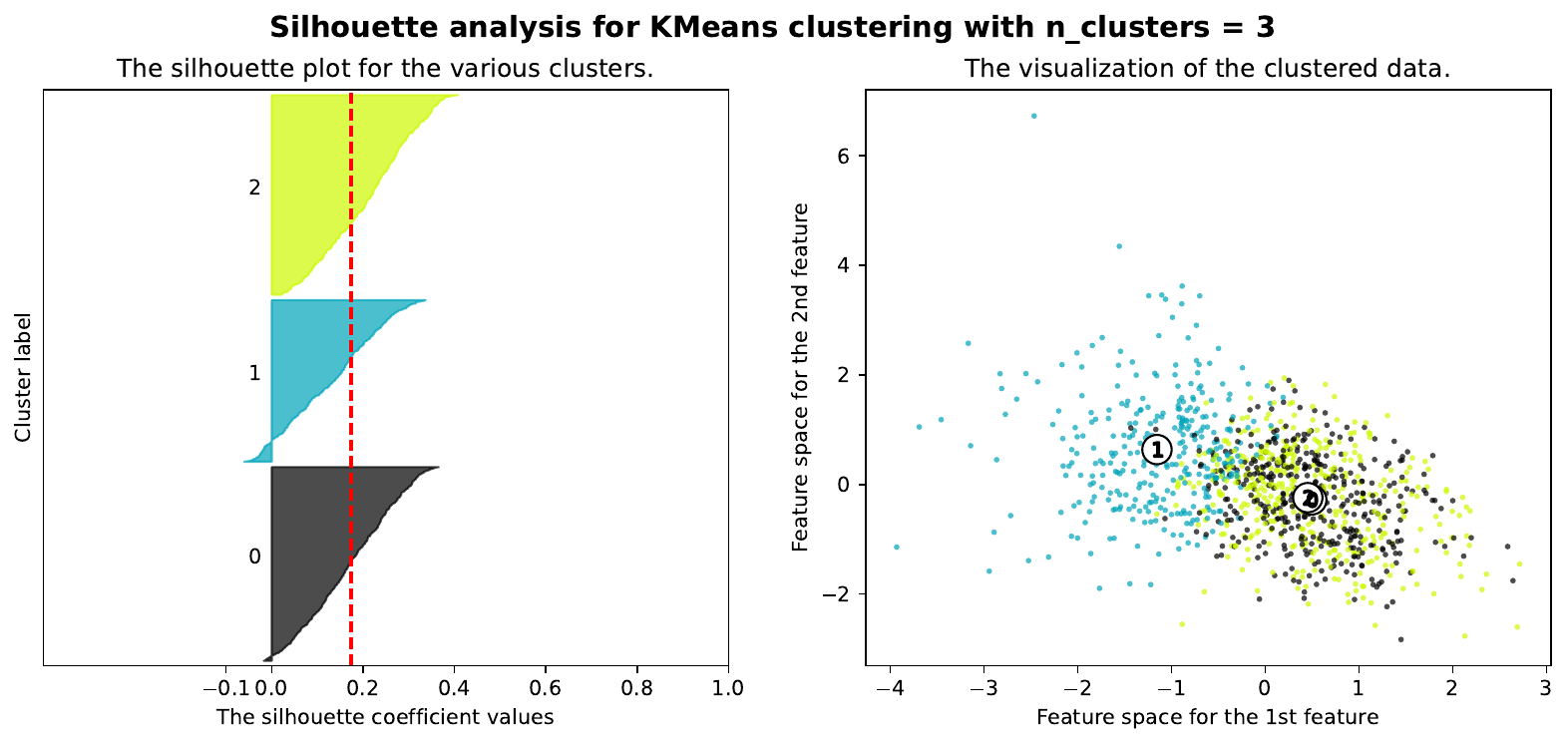} \\
        (a) Elbow Method & (b) Silhouette Method \\[6pt]
         \includegraphics[scale=0.44]{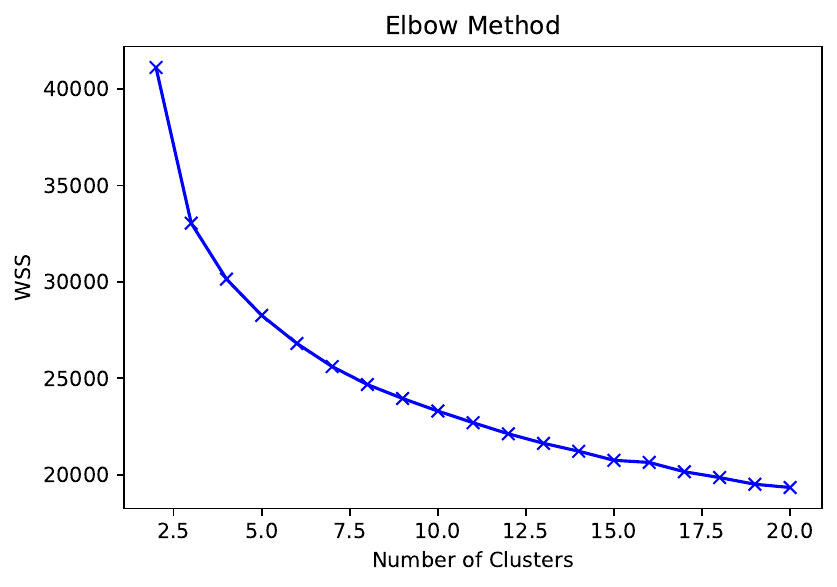} &   \includegraphics[scale=0.38]{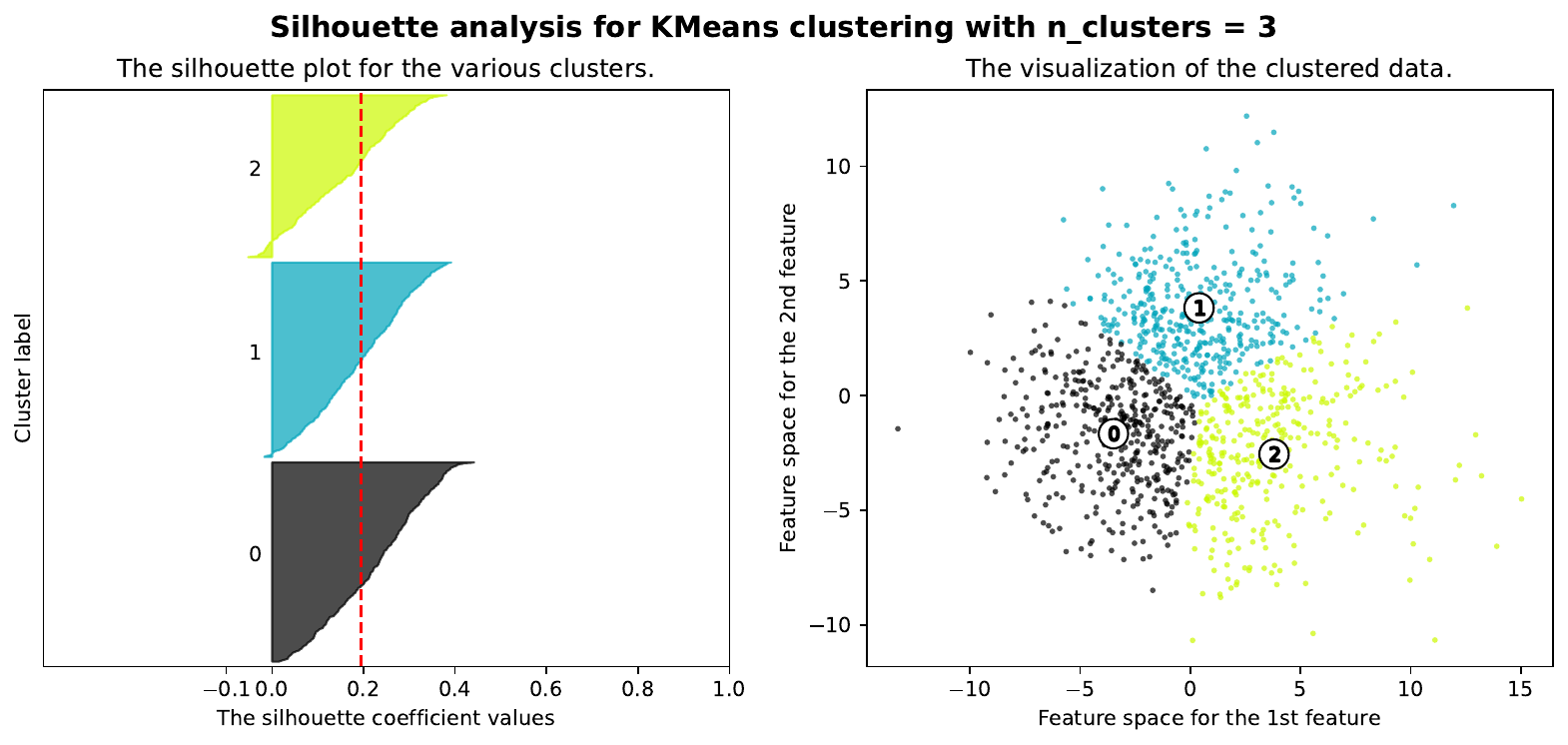} \\
        (c) Elbow Method with PCA & (d) Silhouette Method with PCA \\[6pt]
        \end{tabular}
        \caption{The Elbow and Silhouette methods with a value of $k=3$. Figures \textbf{(a)} and \textbf{(b)} correspond to a dataset without variable selection, while Figures \textbf{(c)} and \textbf{(d)} correspond to the dataset with the use of PCA.}
        \label{fig:cluster}
    \end{figure*}
    
    The $k$-means algorithm was employed for clustering by varying the values of $k$ (number of neighbors) in a range from 2 to 7. To evaluate and analyze the optimal value for $k$, two methods were utilized: the Elbow method and the Silhouette method \cite{rousseeuw1987silhouettes}, as observed in Figure \ref{fig:cluster}. Additionally, the Normalized Mutual Information (NMI) measure \cite{NMI} was used to quantify the information contained in the obtained clusters about the different leagues. However, the results were inconclusive, with very high values for the within-cluster sum of squares in the Elbow method and very low values for the silhouette coefficient and NMI for all tested $k$ values. The best result was achieved with $k=3$, with a silhouette coefficient of 0.173 and NMI approximately at 0.03.
    
    To improve the clustering process of tournaments, the application of Principal Component Analysis (PCA) technique \cite{pca} was tested before applying the $k$-means algorithm. The use of PCA aims to reduce the dimensionality of the data while retaining the most relevant information for analysis and facilitating the identification of patterns. PCA is particularly efficient in datasets with high dimensionality and multicollinearity, which fits this case.
    
\begin{figure}
        \centering
        \includegraphics[scale=0.342]{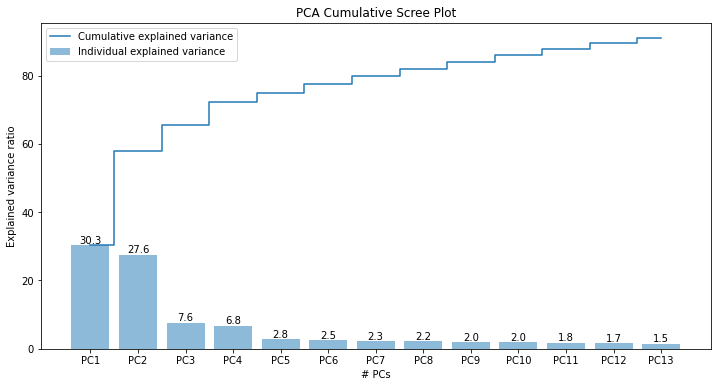}
        \caption{This graph shows the explained variability of each principal component generated by the application of Principal Component Analysis (PCA) technique. The explained variability is expressed in terms of percentage, representing the amount of data variation that is explained by each component. The bar graph indicates the explained variability of each individual component, while the blue line represents the cumulative variability, i.e., the cumulative sum of the explained variability up to the current component.}
        \label{fig:pca_exp}
    \end{figure}

    As observed in Figure \ref{fig:pca_exp}, \textcolor{black}{the first two principal components explain more than half of the variation in the data.} This indicates that they are the most relevant for analysis and contain the most important information for identifying patterns in the data. Therefore, by applying PCA before the k-means algorithm, we efficiently reduce the dimensionality of the data while preserving most of the relevant information.

    A slight improvement in the model was observed, with a silhouette coefficient of 0.194 and NMI of 0.033 for k=3. However, the homogeneity among the clusters persisted, without defining any distinct characteristics among the tournaments.
    
    When testing the algorithm to cluster the matches, it was observed that the obtained clusters exhibited a certain homogeneous distribution among matches from different leagues. This suggests that there are no distinct features among the tournaments in different countries, indicating that playing styles or the structure of passing networks are quite similar among them.
    
    In summary, based in our data, we see that the clustering of tournaments revealed that there are no significant differences in playing styles among the analyzed countries.

\subsection{Complex network results}

    \textcolor{black}{For the construction of the input data for the model, we used the rolling average of the metrics obtained from the last five matches of each team, whether at home or away. Specifically, for each match, we combined the network information from both teams involved in the upcoming game. The dataset was structured such that half of the features represented the metrics of the home team and the other half represented the metrics of the away team.}
    
    \textcolor{black}{For each match, the values were filled with the respective team's data. For example, if Team A was playing at home against Team B, the features related to Team A would include the average metrics from their previous five home matches, and the features related to Team B would include the average metrics from their previous five away matches.}
    
    \textcolor{black}{The target variable for the model was a binary outcome indicating the match result from the perspective of the home team. A value of 1 was assigned if the home team won (and the away team lost), while a value of 0 was assigned if the home team lost (and the away team won).}
    
    \textcolor{black}{This approach ensures that each row in the dataset corresponds to a specific match, with features capturing the recent performance trends of both the home and away teams. By training the model on this dataset, we aimed to predict whether the home team would win or lose the match based on the recent performance metrics of both teams.} 
    
    The test dataset comprised 30\% of the total matches. In the predictive analysis using passing networks, the metrics of accuracy, precision, recall, F1 and AUC were used to evaluate the effectiveness of the model. In this project, more attention was given to accuracy and AUC metrics.

    Two distinct analyses were conducted: one considering a single passing network per team for the entire match and another using a passing network per team for each half of the game (first and second half). The idea of using separate networks for each half was to assess whether passing networks exhibited different patterns in each period of the match, which could potentially improve the prediction of match outcomes.

\begin{figure*}
        \centering
        \includegraphics[scale=0.4]{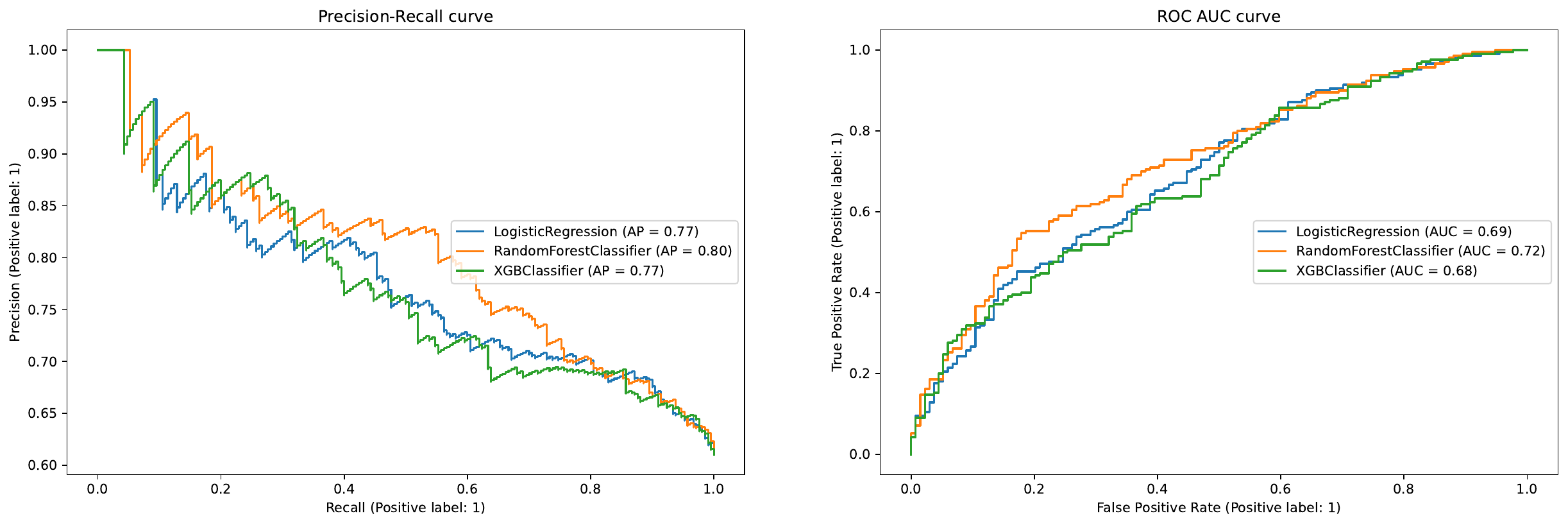}
        \caption{Graphs showing the performance of the different models in relation to the Precision-Recall curve on the left and the ROC-AUC curve on the right.}
        \label{fig:roc_curve}
    \end{figure*}

    \textcolor{black}{The results showed that the analysis using separate passing networks for each half had a slightly better accuracy compared to the analysis using a single passing network for the entire match, reaching an AUC of 0.72, as presented in Figure \ref{fig:roc_curve}, and approximately 75.5\% F1 score in the model, while the entire match network reaches an AUC of 0.70 and 74\%. However, both models performed better than random chance.}

\subsection{Comparing predictive models with and without complex network metrics}

    In this section, the objective is to compare the traditionally used models in this type of prediction, which rely on game statistics, with the hypothesis proposed in the previous section.

    The predictions based on statistics, as well as the models based on passing networks, were also performed using the average of the previous five matches. The following features were used for constructing the model:
    
    \begin{itemize}
        \item Number of goalkeeper saves;
        \item Number of red cards;
        \item Number of yellow cards;
        \item Number of assists;
        \item Number of shots;
        \item Number of opponent shots;
        \item Number of shots on target;
        \item Number of passes;
        \item Number of goals;
        \item Number of opponent goals;
        \item Ball possession;
        \item Accuracy of successful passes;
        \item Accuracy of successful goalkeeper saves;
        \item Accuracy of shots on target.
    \end{itemize}
    
    The results obtained with the predictions based on statistics were very similar to the results obtained with the use of complex network metrics, as presented in Table \ref{tabela:métricas_nets}. This result demonstrates that it is possible to replace traditional predictions based on game statistics with the use of complex network metrics, as both models presented similar results.

    \begin{table}[h]
        \caption{Model evaluation metrics -- Comparing the model based on passing network metrics (upper part) with models based on game statistics (lower part).}
        \label{tabela:métricas_nets}
        \resizebox{\columnwidth}{!}{\begin{tabular}{l|ccccc}
        Model - Nets  & Accuracy & Precision & Recall & F1 & AUC \\ \hline
        Logistic Regression & 67.64\% & 70.42\% & 80.86\% & 75.28\% & 0.75 \\
        Random Forest & 67.93\% & 70.71\% & 80.86\%  & 75.45\% & 0.74    \\
        XG Boost & 66.18\% & 67.82\% & 84.69\% & 75.32\% & 0.69    \\ \hline \hline
        \end{tabular}}
        \resizebox{\columnwidth}{!}{\begin{tabular}{l|ccccc}
        Model - Statistics      & Accuracy & Precision & Recall & F1 & AUC \\ \hline
        Logistic Regression        & 65.12\%      & 64.80\%        & 93.81\%        & 76.65\%    & 0.69    \\
        Random Forest & 66.86\%       & 72.64\%        & 73.33\%        & 72.99\%    & 0.72    \\
        XG Boost       & 66.86\%      & 65.38\%        & 97.14\%        & 78.16\%    & 0.69    \\ \hline
        \end{tabular}}
    \end{table}

\subsection{Combining approaches}

    In this section, a combined analysis of the previous approaches was conducted, once again using separate datasets for each half (first half and second half) and the entire match. The goal was to evaluate whether combining these two datasets could improve the effectiveness of the predictive model.

    The results obtained successfully demonstrated that the combined model, utilizing both statistics features and network metrics, outperformed both approaches used separately. It achieved an accuracy of approximately 71.5\% and an AUC of 0.77, as shown in Table 2.
    
    In an attempt to achieve even better results, techniques such as the removal of highly correlated variables and Principal Component Analysis (PCA) were also tested. These techniques aim to reduce dimensionality and, in some cases, produce more meaningful results for the model. However, this was not observed for the specific model in question. Although it still achieved better results than the individual approaches, it did not surpass the performance of the model containing all variables.
    
    It is important to note that, although dimensionality reduction approaches did not yield significant improvement, the feature combination approach showed promising results. It indicated that the joint use of game statistics and passing network metrics can be an effective strategy in predicting soccer match outcomes.

    \begin{table}[htb]
        \centering
        \caption{Accuracy and AUC compared in the three different models: Based on passing networks, game statistics, and a combination of both. The best result was observed with the Random Forest algorithm using the mixed dataset.}
        \label{tabela:tabela_comparativa}
        \resizebox{\columnwidth}{!}{\begin{tabular}{l||c|c||c|c||c|c}
        Model &  \multicolumn{2}{c||}{Nets}  &  \multicolumn{2}{c||}{Statistics}  &  \multicolumn{2}{c}{Mixed} \\
        \hline
              & Accuracy        & AUC        & Accuracy        & AUC        & Accuracy        & AUC    \\
        \hline
        Logistic Regression        & 67.64\%     & 0.75        & 65.12\%       & 0.69        & 69.39\%     & 0.75    \\
        
        Random Forest & 67.93\%       & 0.74       & 66.86\%        & 0.72    & \textbf{71.44\%}        & \textbf{0.77}    \\
        
        XG Boost       & 66.18\%      & 0.69        & 66.86\%        & 0.69    & 66.18\%    & 0.72    \\
        
        \end{tabular}}
    \end{table}

\subsection{Feature Importance}

    In this section, the permutation importance method for Random Forest from the scikit-learn library \cite{scikit-learn} was used to identify the 20 most important variables in the model, avoiding bias from variables with high cardinality. Additionally, SHAP values were utilized to identify the 20 most impactful features in the model.

    \begin{figure*}
        \centering
        \includegraphics[width=1.5\columnwidth]{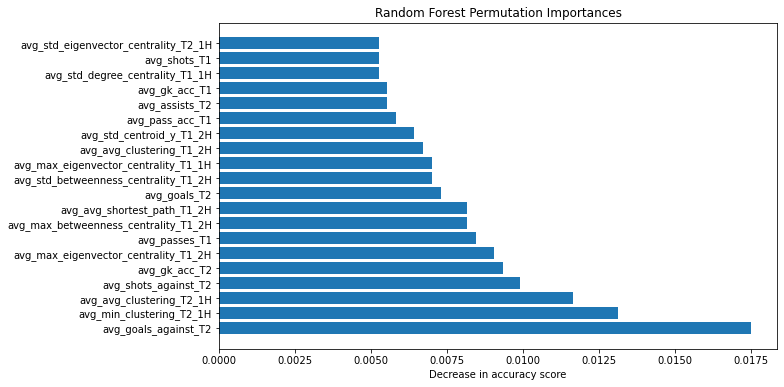}
        \caption{All variable names begin with ``avg'' because they represent the average of the last 5 matches played by the home team (\textit{``T1''}) or away team (\textit{``T2''}). In the second parameter, for network metric variables, the prefixes can be ``min'', \textit{``max''}, \textit{``avg''}, or \textit{``std''}, which correspond to the minimum, maximum, average, or standard deviation of the node metrics in the respective network. Lastly, in the network metrics, the suffixes can be \textit{``1H''} or \textit{``2H''}, which denote the first or second half of the match.}
        \label{fig:feat_importance}
    \end{figure*}
    
    \textcolor{black}{Observing Figure \ref{fig:feat_importance}, the variable that showed the highest importance in the model was the average number of goals scored by the opposing team in away matches (avg\_goals\_against\_T2), followed by the minimum value of the Clustering Coefficient $C_{tr}$ in the network of the visiting team for the first half of the match (avg\_min\_clustering\_T2\_1H). Recall that $C_{tr}$ measures the percentage of triangles (or closed triplets), which in the passing networks reflects the tendency of a player's passes to create interconnected groups or clusters with their teammates.}

    \begin{figure}
        \centering
        \includegraphics[scale=0.29]{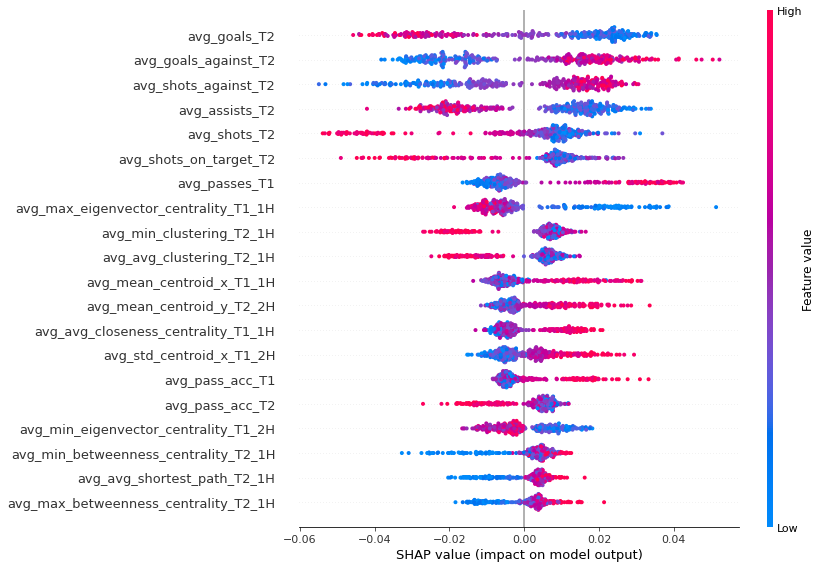}
        \caption{Points with a reddish hue indicate a higher value, while points with a bluish hue indicate a lower value. The further left a point is, the more impact it had on the respective output of the model (in this case, a home team win). The accumulation of points forming ``clouds'' occurs due to the density of values for that variable at a specific SHAP value.}
        \label{fig:shap}
    \end{figure}
    
    \textcolor{black}{In the SHAP plot in Figure \ref{fig:shap}, a wide variety of features was also identified, indicating that both sets of features have an impact on the model. One standout network variable is avg\_max\_eigenvector\_centrality\_T1\_1H, which represents the maximum value of the Eigenvector Centrality for the home team in the first half. This indicates a player who has significant influence with strong connections to other important players in the team, such as a skilled playmaker, for example.}
    
    \textcolor{black}{The results indicate that the most important variables vary between game statistics and network metrics, highlighting the relevance of both in predicting soccer match outcomes.}

    \textcolor{black}{The key variables identified in Figure \ref{fig:feat_importance} do not correspond to those in Figure \ref{fig:shap} because the former displays the top 10 features according to Random Forest Permutation Importance, which measures the decrease in model accuracy when features are shuffled, while the latter shows the top 10 features based on SHAP values, which consider the contribution of each feature to individual predictions. This discrepancy arises due to the different methodologies: Permutation Importance focuses on overall model accuracy, whereas SHAP values highlight the impact of features on individual predictions.}
    
    \textcolor{black}{It is important to note that there were no significant discrepancies between the variables, meaning that no single feature had much greater impact than the others. Instead, the model exhibited a balance in the importance of the variables.}

\subsection{Tournament-based analysis}

    In this section, predictions were made separately for each league in order to identify possible differences among them and compare their predictability. The mixed model, combining network metrics and game statistics, was employed for this purpose.

    The obtained results were quite satisfactory overall, indicating the effectiveness of the approach in predicting soccer match outcomes.
    
    Among the different leagues, the English league emerged as the most predictable, with an accuracy of 80\% using the Random Forest model, while the other leagues ranged from 65\% to 69\%. This suggests that the model successfully captured the specific characteristics and patterns of the English league, resulting in better predictability compared to the other leagues.
    
    It is worth noting that although there were variations in the results across different leagues, the mixed model still exhibited good overall predictive ability, achieving significantly better results than random guessing.

\section{Conclusion}

   In summary, our results indicate that both the model based on standard metrics and the model on network measurements demonstrate similar performance in predicting soccer match outcomes, showing that it is possible to shift from the "traditional" approach to predicting winners and achieve similar results. However, the mixed model that combines both approaches showed an improvement compared to the traditional model, with features from both approaches having significant impact and importance in predicting outcomes. Additionally, these results suggest that teams' strategies change throughout the match, as evidenced by the higher accuracy of the mixed model when considering separate networks for each half compared to a single network for the entire game.
    
    Another important finding is that different leagues do not exhibit significant differences in their characteristics, indicating that models trained on one league can be used to predict outcomes in another league. This discovery can be useful for teams seeking insights about their opponents or for the sports betting market to predict results across different leagues. \textcolor{black}{The fact that the English league emerged as the most predictable suggests that the model successfully captured the specifics characteristics and patterns of those matches, resulting in better predictability. However, the clustering analysis revealed that overall playing styles among the analyzed countries do not exhibit significant differences, which supports the notion that a model trained on one league can generalize to others despite variations in predictability levels.}


    \textcolor{black}{One of the main limitations that may have impacted the quality of the predictions was the size of the available dataset, which was restricted to a single season from five of the most popular leagues in the world, along with a World Cup and a European cup. Additionally, we have focused only on predicting wins and losses of games, without considering other outcomes such as draws. Furthermore, as mentioned earlier, our study is primarily focused on predicting results for the most popular leagues in Europe, which may limit the generalizability of the results to other regions or less popular leagues. The analysis also relied on historical data and did not consider factors such as team changes, coaching changes, player injuries, or other unpredictable circumstances that may occur during a tournament.}



    \textcolor{black}{There are many other scenarios that could be explored with the methods presented here. For instance, one could decrease the temporal granularity, meaning to analyze the data in smaller intervals than a complete half (45 minutes), such as 15-minute intervals for each network. This would provide a better understanding of how game patterns and strategies evolve over time. It would also be interesting to incorporate draws into the analysis. While the choice to focus only on wins and losses was made to simplify the model and achieve better performance, the inclusion of draws can be valuable in understanding how teams perform in situations of equality. Additionally, a promising extension would be to incorporate the information of motifs, which are frequent subgraphs in a network~\cite{gyarmati2014motifs,stone2019network}. The study of motifs can help identify interaction patterns between players that are crucial for team performance. Lastly, one could also employ different machine learning algorithms, such as Support Vector Machines (SVM), Neural Networks, and other boosting-based models like AdaBoost.}

\appendix

\section{Analysis Including Draws}
\label{appendix:appA}

In this section, we revisit our analysis by incorporating matches that resulted in draws into our predictive models. Contrary to our initial approach, where draws were excluded, we sought to understand the impact of including these outcomes on the performance of our models. We conducted experiments using Logistic Regression, Random Forest, and XG Boosting, evaluating their performance with a range of metrics, including accuracy, precision, recall, and F1 score (using the 'macro' approach). 

For this experiment, we retained draws in our dataset, acknowledging their significance in the soccer landscape. The models were trained and evaluated using the same features variables as in our previous analysis, while the target variable were adapted for this experiment to accommodate multiclass classification (win, draw, or loss) instead the binary classification (win or loss) as in our previous analysis.

\begin{table}[ht]
\centering
\caption{Performance Metrics for Models with and Without Draws}
\resizebox{\columnwidth}{!}{\begin{tabular}{lccccccc}
\hline
Model          & \multicolumn{2}{c}{Logistic Regression} & \multicolumn{2}{c}{Random Forest} & \multicolumn{2}{c}{XG Boosting} \\
\hline
Metric                & Without & With & Without & With & Without & With\\
\hline
Accuracy        & 69.39\%           & 49.28\%        & 71.44\%           & 55.03\%        & 66.18\%           & 55.65\%        \\
Precision       & 68.04\%           & 40.91\%        & 69.01\%           & 47.64\%        & 71.13\%           & 47.93\%        \\
Recall          & 65.37\%           & 43.54\%        & 66.49\%           & 46.03\%        & 63.47\%           & 46.61\%        \\
F1 Score        & 65.73\%           & 38.98\%        & 66.92\%           & 41.05\%        & 63.08\%           & 41.89\%        \\
\hline
\end{tabular}}
\label{tabela:tabela_empate}
\end{table}

The inclusion of draws in the dataset had a significant impact on the performance of the prediction models, as shown in Table \ref{tabela:tabela_empate}. While the Random Forest model trained on data without draws initially achieved an accuracy of 71.44\%, the version that incorporated draws reached an accuracy of 55.03\%. This decrease in performance was expected, considering the additional challenge of predicting three possible outcomes (win, draw or loss). However, even with this added complexity, all models outperformed a random baseline of 33\%, highlighting the machine learning model's ability to capture patterns in the data and make predictions beyond random chance.

\section{Simulation}
\label{appendix:appB}        

    In this extended appendix, we provide a detailed explanation of the simulation conducted for the five major soccer leagues: English Premier League, Spanish La Liga, German Bundesliga, French Ligue 1, and Italian Serie A. The simulation was performed using a mixed model with Random Forest, considering the inclusion of draws as a new scenario.
    
    The approach involved leveraging all available games from the remaining leagues as training data to predict the outcomes of matches in the target league. For instance, to predict the French Ligue 1 matches, we utilized all the games from the Italian Serie A, Spanish La Liga, English Premier League, and German Bundesliga. This comprehensive dataset enabled the model to make predictions for each match, classifying them as a win for the home team, a draw, or a win for the away team.
    
    Once the predictions were generated for each league's matches, a simulated championship was constructed, incorporating the respective points earned from the predicted outcomes. The points were allocated based on the traditional scoring system used in soccer leagues, with wins earning three points, draws earning one point, and losses earning zero points. By summing up the points for each team across all simulated matches, the final rankings were determined for each league.

    The simulation results revealed the model's capacity to provide accurate forecasts for match outcomes and simulate league standings across different leagues. Notably, the model successfully predicted the champions in the English Premier League, German Bundesliga, and French Ligue 1 leagues. These findings underscore the predictive potential of the model and its applicability in forecasting soccer championship outcomes in diverse competitive contexts.

    \begin{figure*}
            \begin{tabular}{ccc}
            \includegraphics[scale=0.6]{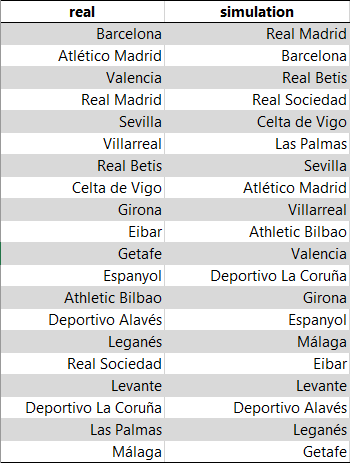} & \includegraphics[scale=0.6]{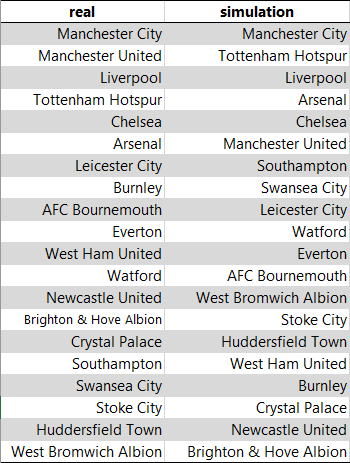} & \includegraphics[scale=0.6]{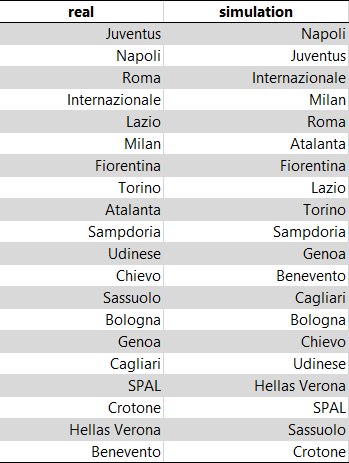}\\
            (a) La Liga (Spain) & (b) Premiere League (England) & (c) Serie A (Italy)\\ 
            \end{tabular}
            \begin{tabular}{cc}
            \includegraphics[scale=0.6]{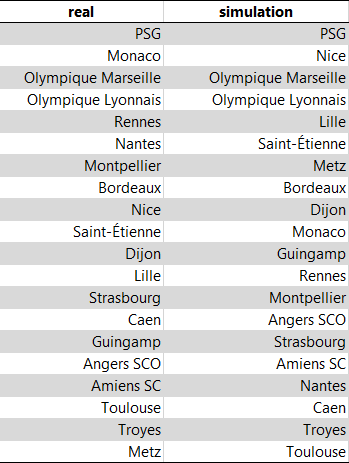} & \includegraphics[scale=0.607]{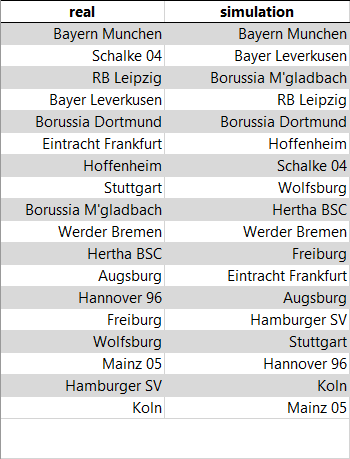}\\
            (d) Ligue 1 (France) & (e) Bundesliga (Germany)\\
            \end{tabular}
            \caption{Simulation results for the five leagues. In the left showing the real results and in the right column the simulated ones. The results are as follows: (a) Spain: 1 exact prediction / 4 predictions with a margin of error of up to 2 positions. (b) England: 3 exact predictions / 8 predictions with a margin of error of up to 2 positions. (c) Italy: 3 exact predictions / 12 predictions with a margin of error of up to 2 positions. (d) France: 5 exact predictions / 10 predictions with a margin of error of up to 2 positions. (e) Germany: 3 exact predictions / 10 predictions with a margin of error of up to 2 positions.}

            \label{fig:simulation}
    \end{figure*}

    The model demonstrated accurate predictions of the real champions in some leagues, such as England, Germany, and France, and also showed promising results with a margin of error of up to 2 positions.

\bibliography{revised_article}

\end{document}